\documentclass[a4paper]{jpconf}
\usepackage{graphicx}
\begin{document}

\title{Characterization of solar-cycle induced frequency shift
of medium- and high-degree acoustic modes}

\author{M. Cristina Rabello-Soares}

\address{Stanford University, Stanford, CA 94305, USA}

\ead{csoares@sun.stanford.edu}

\begin{abstract}

Although it is well known that the solar acoustic mode frequency increases as the solar activity increases, the mechanism behind it is still unknown.
Mode frequencies with $20 < l < 900$ obtained by applying spherical harmonic decomposition to MDI full-disk observations were used.
First,
the dependence of solar acoustic mode frequency with solar activity was examined and evidence of a quadratic relation was found
indicating a saturation effect at high solar activity.
Then, the frequency dependence of 
frequency differences between the activity minimum and maximum
was analyzed. 
The frequency shift scaled by the normalized mode inertia follows a simple power law where the exponent for the $p$ modes
decreases by 37\% for modes with frequency larger than 2.5 mHz.

\end{abstract}

\section{Introduction and Data}

It seems that the responsible mechanism 
for the mode frequencies change with solar activity
is restricted to the outer layers of the Sun.
However, at the moment, there is no general agreement in the precise physical cause.
The frequency variations of medium and high-$l$ modes with 
solar cycle were 
analyzed to help determine its physical origin. 
The solar-radio 10.7-cm daily flux (NGDC/NOAA%\footnote{http://www.ngdc.noaa.gov/stp/solar/solardataservices.html}
) was used as the solar activity proxy.

\begin{figure}[h]
\begin{center}
\includegraphics[width=18pc]{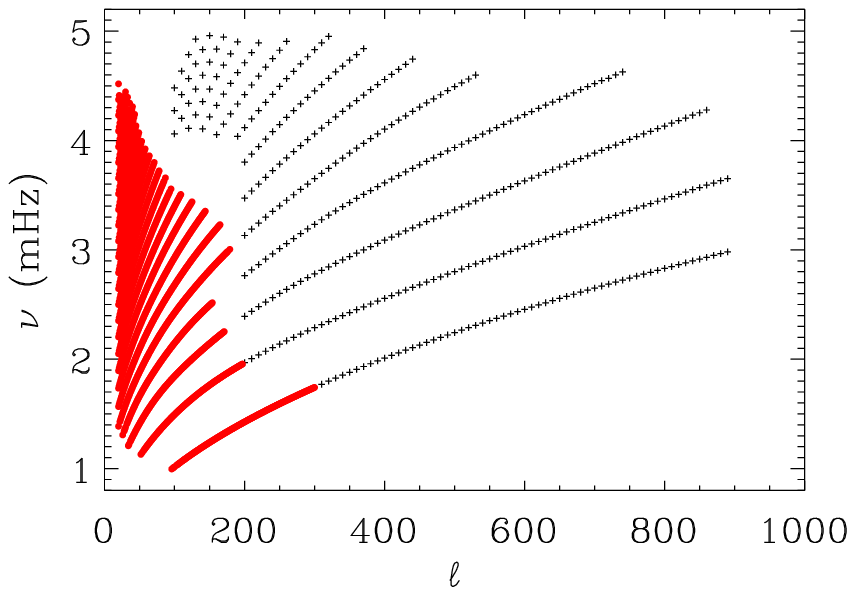}
\includegraphics[width=18pc]{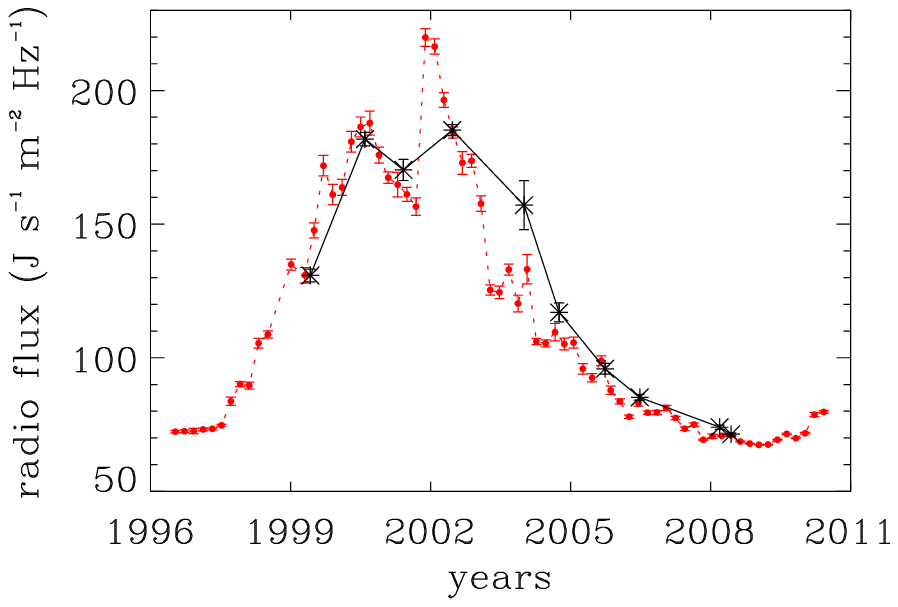}
\end{center}
\caption{\label{lnu} Right panel: Mode coverage. Modes obtained by the medium-$l$ method are in red and by the high-$l$ method in black. Left panel: Mean radio flux for the observational periods in the Dynamics set (in black) and in the Structure set (in red).}
\end{figure}

The 
mode frequencies were obtained by applying spherical harmonic decomposition to MDI full-disk observations
during the Dynamics and Structure observing modes. 
The first one has higher spatial resolution and
is available every year for two or three months of continuous data.
Data for 1999 to 2008 were used.
The second one consists of 72-day time series from early 1996 to April 2010 (Fig.~\ref{lnu}).
They will be called the Dynamics and Structure sets from now on.

The mode frequencies were obtained using two distinct methods.
The medium-$l$ method (\cite{larson} and \cite{schou99}) and
the high-$l$ method (\cite{rabello2008}) which is applied only to the Dynamics set (Fig.~\ref{lnu} left panel).
The second method is used when the spatial leaks of the modes
overlap with the target mode
making it more difficult 
to estimate the mode frequency.

\section{Mode frequency variation with the solar activity cycle}

\begin{figure}[h]
\begin{center}
\includegraphics[width=35pc]{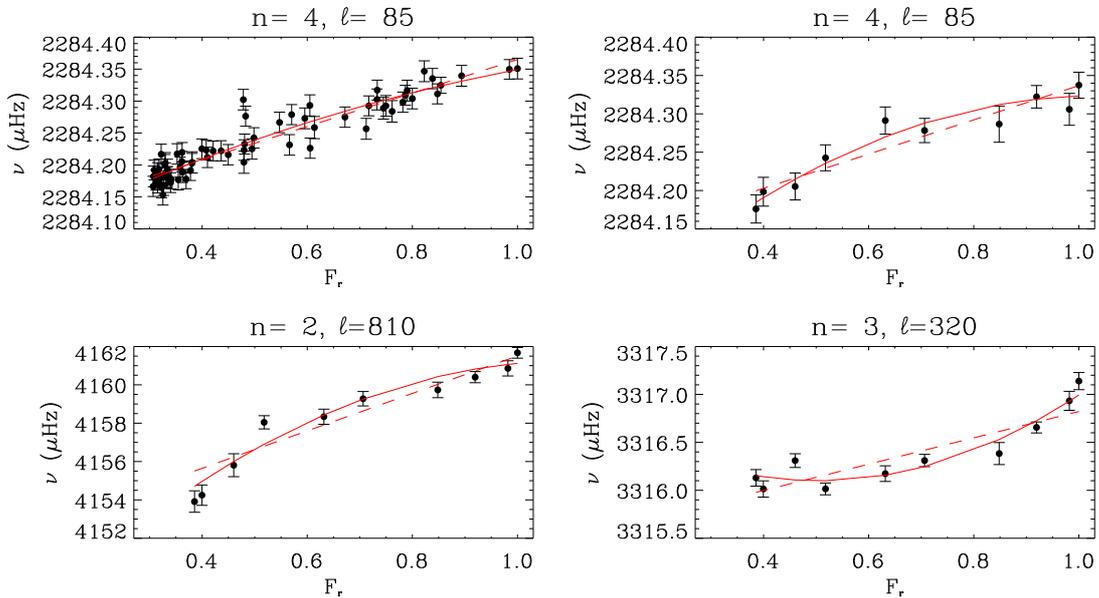}
\end{center}
\caption{\label{fig_quad}Examples of frequency variation with the relative activity index for different modes. The top left panel is using the Structure set and the others using the Dynamics set.}
\end{figure}

There is a very high linear correlation of the mode frequency variation with several solar activity indices.
However, deviations from a simple linear relation has been reported (see, for example, \cite{chaplin2007}). 
In this analysis, there is some indication of a quadratic relation:
\begin{equation}
\nu_q(n,l,F_r) = c_0(n,l) + c_1(n,l) \cdot F_r + c_2(n,l) \cdot F_r^2
\end{equation}
where $F_r$ is the relative activity index defined as the mean solar-radio flux of a given observational period divided by the maximum mean flux of the set of observations (Fig.~\ref{fig_quad}).
The quadratic polynomial seems to indicate a saturation effect at high solar activity. A similar effect has been seen in frequencies at 
activity regions with a large surface magnetic field using ring analysis (\cite{bogart2004}).
The saturation occurs when  $c_1 + 2 \cdot c_2 \cdot F_r^s = 0$, i.e.:
$F_r^s = - c_1 / (2 \cdot c_2)$.

\begin{figure}[h]
\begin{center}
\includegraphics[width=38pc]{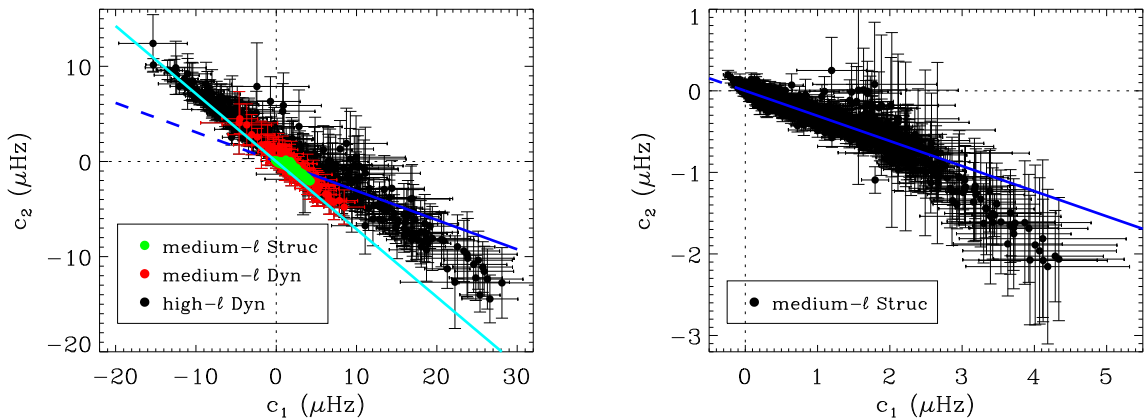}
\end{center}
\caption{\label{fig_c2c1}
Left panel: Dynamics and Structure sets. Right panel: Structure set.
The green circles in the left panel are
the same as the black circles in the right panel. 
The red and black circles in the left panel are for the Dynamics set using the medium- and high-$l$ methods respectively.
The light-blue line is the mean slope of the simulations for the Dynamics set.}
\end{figure}

Fig.~\ref{fig_c2c1} shows the coefficient $c_2(n,l)$ as a function of $c_1(n,l)$.
For the Structure set, the slope of the $c_2-c_1$ linear relation is
$-0.3077 \pm 0.0025$
with a zero y-intercept 
(blue line in both panels of Fig.~\ref{fig_c2c1}),
indicating a saturation at
$(357.2 \pm 2.9) \times 10^{-22}$ J s$^{-1}$ m$^{-2}$ Hz$^{-1}$.
For the Dynamics set,
the slope is $-0.2813 \pm 0.0083$ 
with a very small y-intercept ($-0.0169 \pm 0.0052$ $\mu$Hz). 
For comparison, the linear regression was performed 
on the same $c_1$ range as the Structure set: $0 < c_1 < 4.34$ $\mu$Hz. 
Modes obtained with both medium- and high-$l$ methods were included in the fitting.
Only 12\% (133) of the modes
in the Dynamics set
were obtained with the high-$l$ method.
The saturation occurs at  
$(329.2 \pm 9.7) \times 10^{-22}$ J s$^{-1}$ m$^{-2}$ Hz$^{-1}$.
This is more or less in agreement (differing by $2.8\sigma$) with the
value estimated by the Structure set.
The slope increases as $c_1$ (and $|c_2|$) increases
(Fig.~\ref{fig_c2c1}).
For $c_1 > 4.34$ $\mu$Hz, the slope is $-0.455 \pm 0.028$ and 
the y-intercept $0.50 \pm 0.29$ $\mu$Hz,
giving a saturation at
$(204 \pm 12) \times 10^{-22}$ J s$^{-1}$ m$^{-2}$ Hz$^{-1}$.

\begin{figure}[h]
\includegraphics[width=18pc]{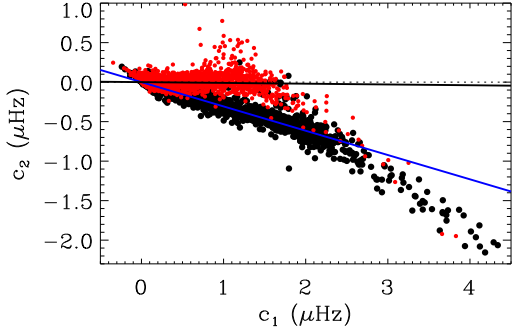}\hspace{1pc}
\begin{minipage}[b]{18pc}
\caption{\label{simul} Coefficient $c_2$ versus $c_1$ for one realization (red circles) and from the observations using the Structure set (black). The black circles are the same as the black circles in the right panel of Fig.~\ref{fig_c2c1}. The black and blue lines are the linear fit to the simulation and to observations respectively. The blue line is the same as the blue line in Fig.~\ref{fig_c2c1}.}
\end{minipage}
\end{figure}

To test the significance of the quadratic relation in the data, the observed frequencies were fitted using a linear relation, $\nu_l(n,l,F_r) = d_0(n,l) + d_1(n,l) \cdot F_r$, and noise was added to the fitted frequencies. To each fitted frequency $\nu_l(n,l,F_r)$, the noise was generated as normally distributed random numbers with the same standard deviation as the error in the observed frequencies. For each $(n,l)$ mode and each realization, a quadratic polynomial was fitted. Fig.~\ref{simul} shows the coefficients obtained for one realization and the ones obtained for the Structure set for comparison.
A straight line was fitted to $c_2(c_1)$ for each realization over the entire range of $c_1$ of each set.
The slope averaged over 1000 realizations is $-0.01437 \pm 0.00011$ and $-0.712222 \pm 0.000060$ for the Structure and Dynamics set respectively.
The first one is represented by the black line in Fig.~\ref{simul} and
the second one by the light-blue line in the left panel of Fig.~\ref{fig_c2c1}.
These slopes are very different from the ones obtained by the corresponding set
indicating that the quadratic relation is not due to noise but to a real signal.

\begin{figure}[h]
\begin{center}
\includegraphics[width=38pc]{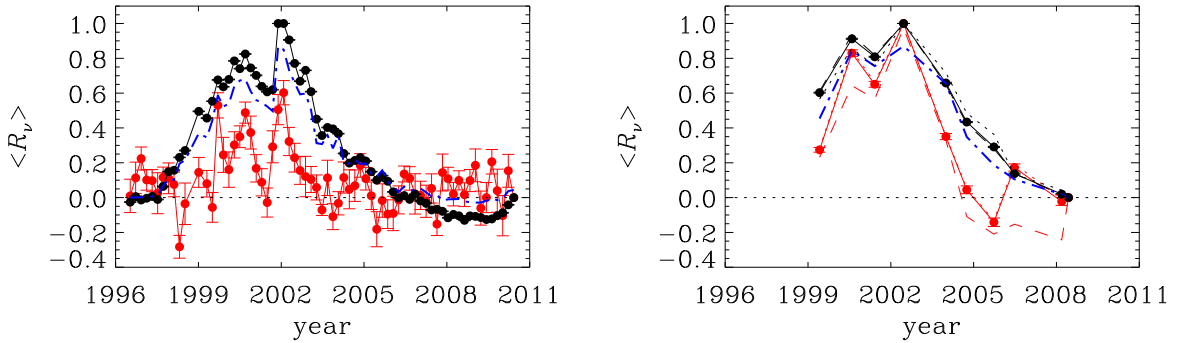}
\end{center}
\caption{\label{fig_c1neg} Left panel: Structure set. Right panel: Dynamics set. Modes with $c_1 > e_{c_1}$ are in black and modes with $c_1 < -e_{c_1}$ in red. The relative activity index is in blue. 
In the right panel, the medium- and high-$l$ modes are given by the dashed and dotted lines respectively.
}
\end{figure}

For the Dynamics set, there are several modes with a negative coefficient $c_1$ (Fig.~\ref{fig_c2c1}). Nine percent (195) of the modes have $c_1 < -e_{c_1}$, where $e_{c_1}$ is the fitting uncertainty. For the Structure set, there are two percent (27) of the modes with $c_1 < -e_{c_1}$. An example of the frequency variation with solar cycle for one of these modes is in the bottom-right panel of Fig.~\ref{fig_quad}.
The frequency of this mode is approximately constant for $F_r < 0.8$.
To see if this is true for all the modes with a negative $c_1$,
the mean relative frequency at a given time $t$ is defined as:
\begin{equation}
\langle R_{\nu}(t)\rangle = \frac{1}{N} ~ \sum_{n,l} ~ \frac{\nu_{n,l}[t] - \nu_{n,l}[min(F_r)]} {\nu_{n,l}[max(F_r)] - \nu_{n,l}[min(F_r)]},
\end{equation}
where $N$ is the number of modes.
$\langle R_{\nu}(t)\rangle$ was estimated over all modes with $c_1 > e_{c_1}$ 
and $c_1 < -e_{c_1}$ given by the black and red circles in Fig.~\ref{fig_c1neg} respectively.
For the Structure set, $N = 1573$ for $c_1 > e_{c_1}$ and $N = 27$ for $c_1 < -e_{c_1}$.
For the Dynamics set, $N = 1272$ for $c_1 > e_{c_1}$ and $N = 60$ for $c_1 < -e_{c_1}$. 
Only modes with a Spearman's correlation coefficient larger than 0.75 
were used in the average of the Dynamics set to decrease the scatter.
Modes with a negative $c_1$ seem to be less sensitive to the solar cycle, more so away from the solar maximum.
Most of these modes have a small $\nu/L$, between 3 and 25 $\mu$Hz.
In the Dynamics set, most of them that are $p$ modes have $3 < \nu < 3.6$ mHz and $l > 200$ while the $f$ modes have $l > 600$.
In the Structure set, most that are $p$ modes have $\nu < 1.6$ mHz and $l < 170$ while the $f$ modes have $100 < l < 140$.
It is not clear why these modes are less affected by the solar cycle.
If this is proved to be true and not an artifact of the data, it will help in the understanding of the mechanism or mechanisms responsible for the mode frequency variation with solar activity.

From Fig.~\ref{fig_c1neg} (left panel), 
the minimum between solar cycle 23 and 24 happened around August 2008.
The frequencies were at a minimum from November 2007 to August 2009.
As pointed out by other authors (see \cite{jain2010} and references within),
the mode frequencies were smaller than the previous minimum.
The frequency averaged over all medium-$l$ modes is $\sim 10\%$ 
of the minimum-to-maximum variation
smaller than in the previous minimum:
$\langle R_{\nu}($July 1996$)\rangle - \langle R_{\nu}($August 2008$)\rangle \approx 0.1$.

\section{The minimum-to-maximum frequency shift}

Here, the minimum-to-maximum frequency shift
$\delta\nu(n,l)$ is defined as the difference between the fitted frequency given by Eq.~(1) at the maximum and minimum solar activity.
The minimum and maximum activity was chosen to be those of the Dynamics set, which are around April 2008 and May 2002 respectively (Fig.~\ref{lnu} right panel). 
The differences in $\delta\nu(n,l)$ using a linear regression instead of Eq.~(1) are small.
The standard deviation of the differences is 0.0031 $\mu$Hz and 0.25 $\mu$Hz for the Structure and Dynamics set respectively.
Most of the difference is due to the high-$l$ modes in the Dynamics set.
The effect of these differences in the estimated $\alpha$ and $\gamma$ coefficients described below is in their 4-th significative digit or higher.

\begin{figure}[h]
\begin{center}
\includegraphics[width=38pc]{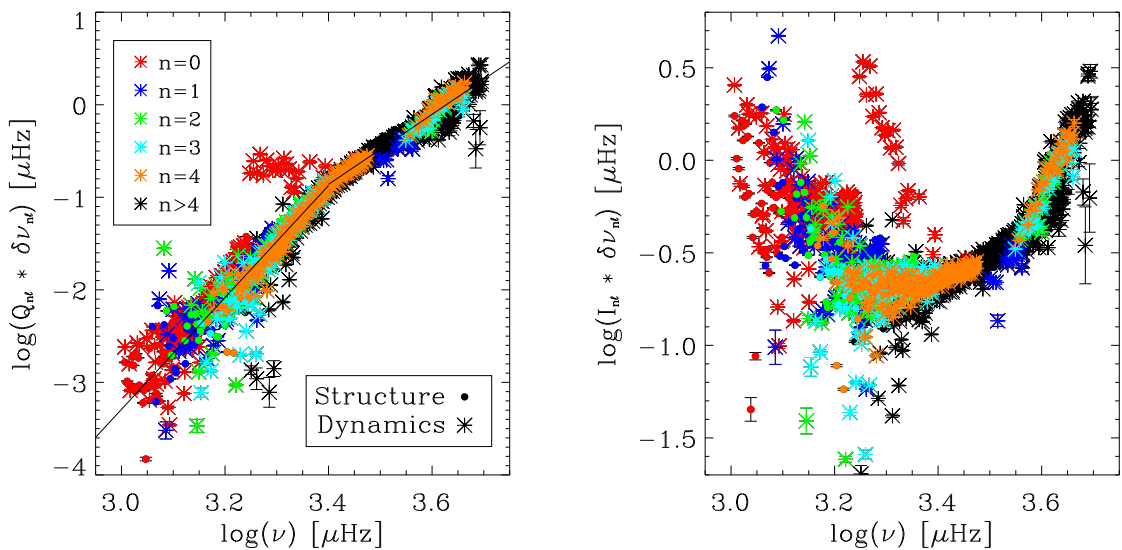}
\end{center} 
\caption{\label{delnu_comp} Only modes with $c_1 > 0$ are plotted. 
%Modes with $n > 4$ are in black.
}
\end{figure}

Most of, if not all, the perturbation to the structure of the Sun associated with the solar cycle is believed to be located in a thin layer near the solar surface which leads to a frequency perturbation roughly proportional to the inverse of the mode inertia $I_{n,l}$:
$\delta\nu_{n,l} \propto \nu_{n,l}^{\alpha}/I_{n,l}$.
The value of $\alpha$ depends upon the physical mechanism responsible for affecting the mode frequencies during the solar cycle. 
Some of the predicted values are -1, 1 and 3 (see \cite{gough1990} and references within).
Fig.~\ref{delnu_comp} (right panel) shows $\delta\nu(n,l)$ 
scaled by $I_{n,l}$ for both sets. Only modes with $c_1 > 0$ were plotted.
Fitting only the $p$ modes, $\alpha$ varies from approximately -4 for $\nu < 1.6$ mHz, to 0. for $1.66 < \nu < 2.5$ mHz, to $1.43772 \pm 0.00018$ for $2.4 < \nu < 3.55$ mHz, to $6.417 \pm 0.023$ for $\nu > 4$ mHz.
The variation of $\alpha$ indicates a more complex mechanism than anticipated.

Recently, \cite{rabello2008}
observed that scaling the frequency shift with the mode inertia normalized by the inertia of a radial mode of the same frequency ($Q_{n,l}$)
also follows a simple power law:
$\delta\nu_{n,l} \propto \nu_{n,l}^{\gamma}/Q_{n,l}$,
but with a constant exponent.
Extending this previous analysis to include data from the end of solar cycle 23 
gives similar results
(Fig.~\ref{delnu_comp} left panel).
However, there is a clear change in the coefficient $\gamma$ around 2.5 mHz.
For the $p$ modes, 
$\gamma_p = 6.04466 \pm 0.00030$ for $\nu < 2.5$ mHz and
$\gamma_p = 3.78657 \pm 0.00019$ for $\nu > 2.5$ mHz.
In the previous analysis, it was estimated $\gamma_p = 3.60 \pm 0.01$ at all frequencies and it was argued that around 2.3 mHz there was a step in the $p$-mode frequency. This analysis show clearly that there in fact a change in $\gamma_p$.

The $p$ modes obtained by the high-$l$ method 
show a slightly different slope than those obtained by the medium-$l$ method. 
In Fig.~\ref{delnu_comp},
the $p$ modes obtained by the high-$l$ method are concentrated at $\log[\nu$($\mu$Hz)$] > 3.55$ where they account for 70\% of the modes (including the low $n$ modes indicated by the colored stars).
Although an artifact from the analysis can not be discarded due to the difficulty in estimating unbiased mode frequencies at high-$l$ and/or high-frequency, it could be an indication of a degree dependence in the scaled frequency shift.

\begin{figure}[h]
\begin{center}
\includegraphics[width=38pc]{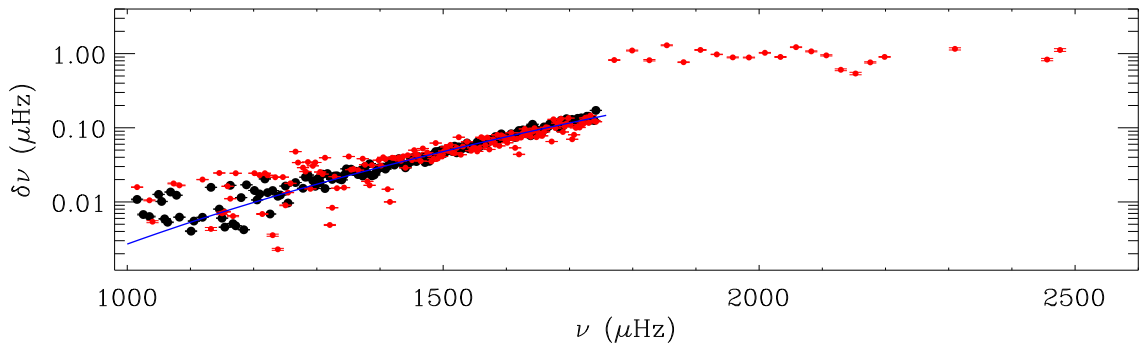}
\end{center}
\caption{\label{fmodes} Only f modes. In red are the Dynamics set and in black the Structure set. The high-$l$ modes are those with $\nu > 1760$ $\mu$Hz. The blue line is for $\gamma_f = 7.0880 \pm 0.0019$.}
\end{figure}

It is expected that the $f$ modes are affected by the solar cycle in a different way than the $p$ modes since they have very different properties.
Fig.~\ref{fmodes} shows the $f$-mode frequency shift.
The high-$l$ $f$ modes in the Dynamics set has a distinctive behavior (red stars in Fig.~\ref{delnu_comp} with $3.25 < \log[\nu$($\mu$Hz)$] < 3.4$).
Fitting only the modes obtained by the medium-$l$ method, 
$\gamma_f = 7.0880 \pm 0.0019$ which is 17\% larger than $\gamma_p$ at the corresponding frequency range.

\section{Summary}

There is some evidence of a quadratic relation of the frequency shifts with the solar-radio flux indicating a saturation effect at high solar activity. The saturation occurs at $\sim350 \times 10^{-22}$ J s$^{-1}$ m$^{-2}$ Hz$^{-1}$ and decreases as $c_1$ increases.

The minimum-to-maximum frequency shift scaled by the mode inertia is proportional to $\nu^{\alpha}$ where 
$\alpha_p$ increases from -4 at low frequency to 6.4 at high frequency for the $p$ modes.
Scaling by the normalized mode inertia, 
there is a sharp change in the $p$-mode exponent at 2.5 mHz.

\ack
The frequencies obtained by the medium-$l$ method 
were calculated by J. Schou and T. Larson.

\end{document}